\documentclass{article}
\usepackage{spconf,amsmath,graphicx}
\usepackage{amsfonts}
\usepackage{gensymb}
\usepackage{algorithmic}
\usepackage{graphicx}
\usepackage{textcomp}
\usepackage{xcolor}
\usepackage{hyperref}
\usepackage{subfigure}
\usepackage[bottom]{footmisc}

\title{VSEGAN: Visual Speech Enhancement Generative Adversarial Network}
\name{Xinmeng Xu$^{1,2}$, Yang Wang$^1$, Dongxiang Xu$^1$, Yiyuan Peng$^1$, Cong Zhang$^1$, Jie Jia$^1$, Binbin Chen$^1$}
\address{$^1$vivo AI Lab, P.R. China\\
  $^2$E.E. Engineering, Trinity College Dublin, Ireland}
\begin{document}
%
\maketitle
\begin{abstract}
Speech enhancement is an essential task of improving speech quality in noise scenario. Several state-of-the-art approaches have introduced visual information for speech enhancement, since the visual aspect of speech is essentially unaffected by acoustic environment. This paper proposes a novel framework that involves visual information for speech enhancement, by incorporating a Generative Adversarial Network (GAN). In particular, the proposed visual speech enhancement GAN consists of two networks trained in adversarial manner, i) a generator that adopts multi-layer feature fusion convolution network to enhance input noisy speech, and ii) a discriminator that attempts to minimize the discrepancy between the distributions of the clean speech signal and enhanced speech signal. Experiment results demonstrated superior performance of the proposed model against several state-of-the-art models. 
\end{abstract}
\section{Introduction}
\label{sec:intro}

Speech processing systems are used in a wide variety of applications such as speech recognition, speech coding, and hearing aids. These systems have best performance under the condition that noise interference are absent. Consequently, speech enhancement is essential to improve the performance of these systems in noisy background \cite{loizou2013speech}. Speech enhancement is a kind of algorithm that can be used to improve the quality and intelligibility of noisy speech, decrease the hearing fatigue, and improve the performance of many speech processing systems. 

Conventional speech enhancement algorithms are mainly based on signal processing techniques, e.g., by using speech signal characteristics of a known speaker, which include spectral subtraction \cite{martin1994spectral}, signal subspace \cite{ephraim1995signal}, Wiener filter \cite{lim1978all}, and model-based statistical algorithms \cite{dendrinos1991speech}. Various deep learning networks architectures, such as fully connected network, Convolution Neural Networks (CNNs), Recurrent Neural Networks (RNNs), have been demonstrated to notably improve speech enhancement capabilities than that of conventional approaches. Although deep learning approaches make noisy speech signal more audible, there are some remaining deficiencies in restoring intelligibility. 

Speech enhancement is inherently multimodal, where visual cues help to understand speech better. The correlation between the visible proprieties of articulatory organs, e.g., lips, teeth, tongue, and speech reception has been previously shown in numerous behavioural studies \cite{sumby1954visual}. Similarly, a large number of previous works have been developed for visual speech enhancement, which based on signal processing techniques and machine learning algorithms \cite{wang2005video}. Not surprisingly, visual speech enhancement has been recently addressed in the framework of DNNs, a fully connected network was used to jointly process audio and visual inputs to perform speech enhancement \cite{hou2016audio}. The fully connected architecture cannot effectively process visual information, which caused the audio-visual speech enhancement system slightly better than its audio-only speech enhancement counterpart. In addition,
there is a model which feed the video frames into a trained speech generation network, and predict clean speech from noisy input \cite{gabbay2018seeing}, which has shown more obvious improvement when compared with the previous approaches. 

The Generative Adversarial Network (GAN) consists of a generator network and a discriminator network that play a min-max game between each other, and GAN have been explored for speech enhancement, SEGAN \cite{inproceedings} is the first approach to apply GAN to speech enhancement model. This paper proposes a Visual Speech Enhancement Generative Adversarial Network (VSEGAN) that enhances noisy speech using visual information under GAN architecture.

The rest of article is organized as follows: Section 2 presents the proposed method in detail. Section 3 introduces the experimental setup. Experiment results are discussed in Section 4, and a conclusion is summarized in Section 5.

\begin{figure*}[t]
  \centering
  \includegraphics[width=0.7\linewidth]{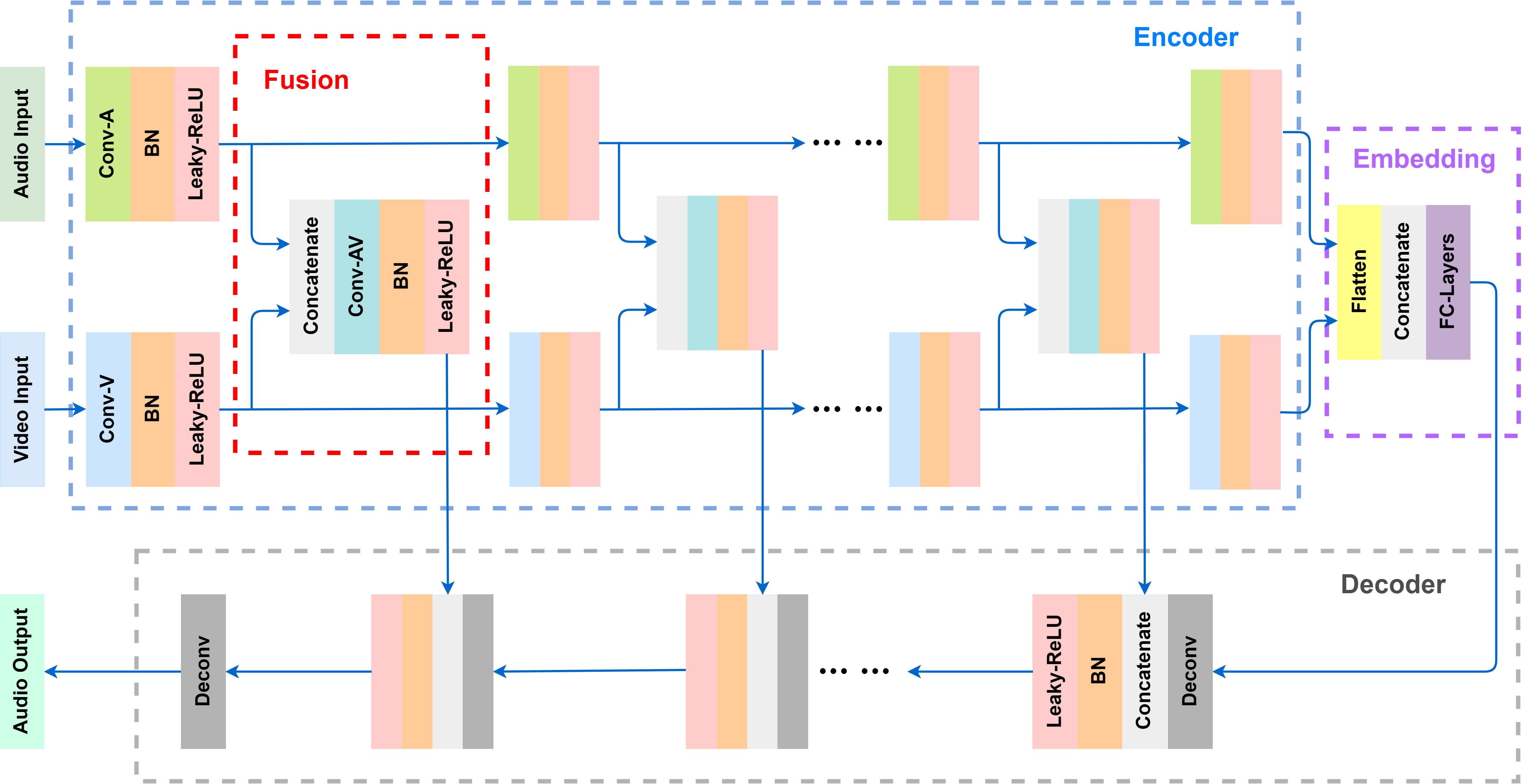}
  \caption{Network architecture of generator. Conv-A, Conv-V, Conv-AV, BN, and Deconv denote convolution of audio encoder, convolution of video encoder, convolution of audio-visual fusion, batch normalization, and transposed convolution.}
  \label{fig:generator}
\end{figure*}

\section{Model Architecture}

\subsection{Generative Adversarial Network}
GAN is comprised of generator (G) and discriminator (D). The function of G is to map a noisy vector $\textbf{x}$ from a given prior distribution $\mathcal{X}$ to an output sample $\textbf{y}$ from the distribution $\mathcal{Y}$ of training data. D is a binary classifier network, which determines whether its input is real or fake. The generated samples coming from $\mathcal{Y}$, are classified as real, whereas the samples coming from G, are classified as fake. The learning process can be regarded as a minimax game between G and D, and can be expressed by:
\begin{equation}
    \begin{aligned}
        \mathop{\rm min}\limits_{G} \mathop{\rm max}\limits_{D} V(D,G)&=\mathbb{E}_{\textbf{y}\sim p_{\textbf{y}}(\textbf{y})}[\log(D(\textbf{y}))]\\
      &+\mathbb{E}_{\textbf{x}\sim p_{\textbf{x}}(\textbf{x})}[\log(1-D(G(\textbf{x})))]
    \end{aligned}
\end{equation}

Training procedure for GAN can be concluded the repetition of following three steps:
\begin{itemize}
    \item[] Step 1: D back-props a batch of real samples $\textbf{y}$.
    \item[] Step 2: Freeze the parameters of G, and D back-props a batch of fake samples that generated from G.
    \item[] Step 3: Freeze the parameters of D, and G back-props to make D misclassify.
\end{itemize}

The regression task generally works with a conditioned version of GAN \cite{mirza2014conditional}, in which some extra information, involve in a vector $\textbf{y}_{c}$, is provided along with the noisy vector $\textbf{x}$ at the input of G. In that case, the cost function of D is expressed as following:
\begin{equation}
    \begin{aligned}
        \mathop{\rm min}\limits_{G}\mathop{\rm max}\limits_{D} &V(D,G)=\mathbb{E}_{\textbf{y},\textbf{y}_{c}\sim p_{\textbf{y}}(\textbf{y},\textbf{y}_c)}[\log(D(\textbf{y},\textbf{y}_c))]\\
      & +\mathbb{E}_{\textbf{x}\sim p_{\textbf{x}}(\textbf{x}),\textbf{y}_{c}\sim p_{\textbf{y}}(\textbf{y}_c)}[\log(1-D(G(\textbf{x},\textbf{y}_{c}),\textbf{y}_{c}))]
    \end{aligned}
\end{equation}

However, Eq. (2) are suffered from vanishing gradients due to the sigmoid cross-entropy loss function \cite{aaa}. To tackle this problem, least-squares GAN approach \cite{mao2017least} substitutes cross-entropy loss to the mean-squares function with binary coding, as given in Eq. (3) and Eq. (4).   
\begin{equation}
    \begin{aligned}
        \mathop{\rm max}\limits_{D} &  V(D)=\frac{1}{2}\mathbb{E}_{\textbf{y},\textbf{y}_{c}\sim p_{\textbf{y}}(\textbf{y},\textbf{y}_c)}[\log(D(\textbf{y},\textbf{y}_c)-1)^2]\\
      &+\frac{1}{2}\mathbb{E}_{\textbf{x}\sim p_{\textbf{x}}(\textbf{x}),\textbf{y}_{c}\sim p_{\textbf{y}}(\textbf{y}_c)}[\log(1-D(G(\textbf{x},\textbf{y}_{c}),\textbf{y}_{c}))^2]
    \end{aligned}
\end{equation}

\begin{equation}
    \begin{aligned}
        \mathop{\rm min}\limits_{G} &  V(G)=\frac{1}{2}\mathbb{E}_{\textbf{x}\sim p_{\textbf{x}}(\textbf{x}),\textbf{y}_{c}\sim p_{\textbf{y}}(\textbf{y}_c)}[\log(D(G(\textbf{x},\textbf{y}_{c}),\textbf{y}_{c})-1)^2]
    \end{aligned}
\end{equation}

\subsection{Visual Speech Enhancement GAN}

\begin{figure*}
    \includegraphics[width=0.73\linewidth]{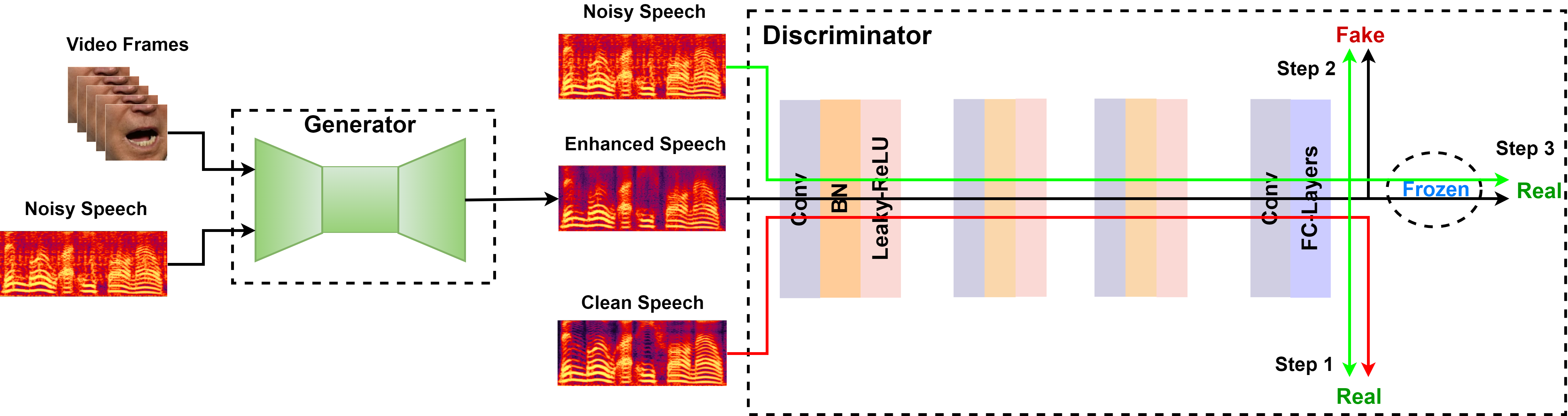}
    \centering
    \caption{Network architecture of discriminator, and GAN training procedure.}
    \label{fig:discriminator}
\end{figure*}

\renewcommand{\arraystretch}{1.0}
\begin{table*}[]
\caption{Detailed architecture of the VSEGAN generator encoders. Conv1 denotes the first convolution layer of the VSEGAN generator encoder part.}
\begin{center}
\begin{tabular}{c|cccccccccc}
\hline
                 & Conv1     & Conv2     & Conv3     & Conv4     & Conv5     & Conv6     & Conv7     & Conv8     & Conv9     & Conv10    \\ \hline
Num Filters      &   64      &   64      &   128     &   128     &  256      &  256      &  512      &  512      &  1024     &  1024      \\
Filter Size      & (5, 5)    & (4, 4)    & (4, 4)    & (4, 4)    & (2, 2)    & (2, 2)    & (2, 2)    & (2, 2)    & (2, 2)    & (2, 2)    \\
Stride(audio)    & (2, 2)    & (1, 1)    & (2, 2)    & (1, 1)    & (2, 1)    & (1, 1)    & (2, 1)    & (1, 1)    & (1, 5)    & (1, 1)    \\
MaxPool(video)   & (2, 4)    & (1, 2)    & (2, 2)    & (1, 1)    & (2, 1)    & (1, 1)    & (2, 1)    & (1, 1)    & (1, 5)    & (1, 1)  \\ \hline
\end{tabular}
\end{center}
\end{table*}
The G network of VSEGAN performs enhancement, where its inputs are noisy speech $\tilde{\textbf{y}}$ and video frames $\textbf{v}$, and its output is the enhanced speech $\widehat{\textbf{y}}=G(\tilde{\textbf{y}}, \textbf{v})$. The G network follows an encoder-decoder scheme, and consist of encoder part, fusion part, embedding part, and decoder part. The architecture of G network is shown in Figure~\ref{fig:generator}.

Encoder part of G network involves audio encoder and video encoder. The audio encoder is designed as a CNN taking spectrogram as input, and each layer of an audio encoder is followed by strided convolutional layer, batch normalization, and Leaky-ReLU for non-linearity. The video encoder is used to process the input face embedding through a number of max-pooling convolutional layers followed by batch normalization, and Leaky-ReLU. In the G network, the dimension of visual feature vector after convolution layer has to be the same as the corresponding audio feature vector, since both vectors take at every encoder layer is through a fusion part in encoding stage. The audio decoder is reversed in the audio encoder part by deconvolutions, followed again by batch normalization and Leaky-ReLU.

Fusion part designates a merged dimension to implement fusion, and the audio and video streams take the concatenation operation and are through several strided convolution layer, followed by batch normalization, and Leaky-ReLU. Embedding part consists of three parts: 1) flatten audio and visual steams, 2) concatenate flattened audio and visual streams together, 3) feed concatenated feature vector into several fully-connected layers. The output of fusion part in each layer is fed to the corresponding decoder layer. Embedding part is a bottleneck, which applied deeper feature fusion strategy, but with a larger computation expense. The architecture of G network avoids that many low level details could be lost to reconstruct the speech waveform properly, if all information are forced to flow through the compression bottleneck.

The D network of VSEGAN has the same structure with SERGAN \cite{baby2019sergan}, as shown in Figure~\ref{fig:discriminator}. The D can be seen as a kind of loss function, which transmits the classified information (real or fake) to G, i.e., G can predict waveform towards the realistic distribution, and getting rid of the noisy signals labeled to be fake. In addition, previous approaches \cite{isola2017image} demonstrated that using $L_1$ norm as an additional component is beneficial to the loss of G, and $L_1$ norm which performs better than $L_2$ norm to minimize the distance between enhanced speech and target speech \cite{pandey2018adversarial}. Therefore, the G loss is modified as:

\begin{equation}
    \begin{aligned}
        \mathop{\rm min}\limits_{G}V(G) &=\frac{1}{2}\mathbb{E}_{\textbf{x}\sim p_{\textbf{x}}(\textbf{x}),\tilde{\textbf{y}}\sim p_{\textbf{y}}(\tilde{\textbf{y}})}[(D(G(\textbf{x},(\textbf{v}, \tilde{\textbf{y}})),\tilde{\textbf{y}})\\
        &-1)^2]+\lambda||G(\textbf{x},(\textbf{v}, \tilde{\textbf{y}}))-\textbf{y}||_1
     \end{aligned}
\end{equation}
where $\lambda$ is a hyper-parameter to control the magnitude of the $L_1$ norm.

\section{Experiment setup}
\subsection{Datasets}
The model is trained on two datasets: the first is the GRID \cite{cooke2006audio} which consist of video recordings where 18 male speakers and 16 female speakers pronounce 1000 sentences each; the second is TCD-TIMIT \cite{c}, which consist of 32 male speakers and 30 female speakers with around 200 videos each.

The noise signals are collected from the real world and categorized into 12 types: room, car, instrument, engine, train, talker speaking, air-brake, water, street, mic-noise, ring-bell, and music. At every iteration of training, a random attenuation of the noise interference in the range of [-15, 0] dB is applied as  a data augmentation scheme. This augmentation was done  to make the network robust against various SNRs.

\subsection{Training and Network Parameters}
The video representation is extracted from input video and is resampled to 25 frames per seconds. Each video is divided into non-overlapping segments of 5 consecutive frames. The audio representation is the transformed magnitude spectrograms in the log Mel-domain with 80 Mel frequency bands from 0 to 8 kHz, using a Hanning window of length 640 bins (40 milliseconds), and hop size of 160 bins (10 milliseconds). The whole spectrograms are sliced into pieces of duration of 200 milliseconds corresponding to the length of 5 video frames.
 
The proposed VSEGAN has 10 convolutional layers for each encoder and decoder of generator, and the details of audio and visual encoders are described in Table 1, and a Conv-A or a Conv-V in Figure 1 comprise of two convolution layers in Table 1.

The model is trained with ADAM optimizer for 70 epochs with learning rate of $10^{-4}$, and batch size of 8, and the hyper parameter $\lambda$ of loss function in Eq. (5) is set to 100.

\renewcommand{\arraystretch}{1.0}
\begin{table}[]
\caption{Performance of trained networks}
\centering
\begin{tabular}{c|cccc}
\hline
Test SNR           & \multicolumn{2}{c}{-5 dB} & \multicolumn{2}{c}{0 dB} \\ \hline
Evaluation Metrics & STOI         & PESQ        & STOI        & PESQ        \\ \hline
Noisy              & 51.4         & 1.03        & 62.6        & 1.24        \\ 
SEGAN              & 63.4         & 1.97        & 77.3        & 2.21        \\ 
Baseline           & 81.3         & 2.35        & 87.9        & 2.94        \\ 
VSEGAN             & \textbf{86.8}         & \textbf{2.88 }       & \textbf{89.8}        & \textbf{3.10}        \\ \hline
\end{tabular}
\end{table}

\section{Results}
The performance of VSEGAN is evaluated with the following metrics: Perceptual Evaluation of Speech Quality (PESQ), and Short Term Objective Intelligibility (STOI). In addition, there are three networks have trained for comparison:
\begin{itemize}
    \item \textbf{SEGAN} \cite{inproceedings}: An audio-only speech enhancement generative adversarial network.
    \item \textbf{Baseline} \cite{ref39}: A baseline work of visual speech enhancement.
    \item \textbf{VSEGAN}: the proposed model, visual speech enhancement generative adversarial network.
\end{itemize}

\renewcommand{\arraystretch}{1.0}
\begin{table}[]
\caption{Performance comparison of VSEGAN with state-of-the-art result on GRID}
\centering
\begin{tabular}{c|cccc}
\hline
Test SNR           & -5 dB        & 0 dB       & -5 dB         & 0 dB          \\ \hline
Evaluation Metrics & \multicolumn{2}{c}{PESQ} & \multicolumn{2}{c}{STOI(\%)} \\ \hline
L2L                & 2.61         & 2.92       & 85.89         & 88.96         \\
OVA                & 2.69         & 3.00       & 86.17         & 89.75         \\
AV(SE)$^2$             & -            & 2.98       & 86.06         & 89.44         \\
VSEGAN             & 2.88         & 3.10       & 86.84         & 89.83         \\ \hline
\end{tabular}
\end{table}

\begin{figure}[htbp]
	\centering
	\includegraphics[width=0.22\textwidth]{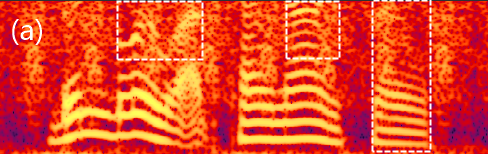} 
	\includegraphics[width=0.22\textwidth]{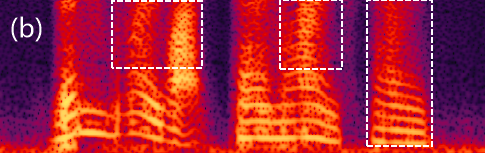}\\
	\includegraphics[width=0.22\textwidth]{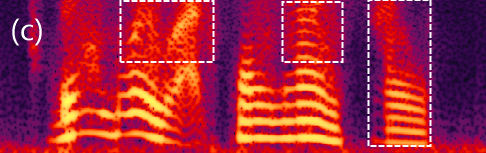}
	\includegraphics[width=0.22\textwidth]{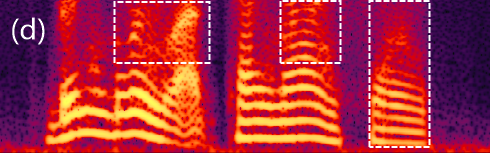}
	\caption{Example of input and enhanced spectra from an example speech utterance. (a) Noisy speech under the condition of noise at 0 dB. (b) Enhanced speech generated by baseline work. (c) Enhanced speech generated by Generator. (d) Enhanced speech generated by VSEGAN.}
	\label{fig:spe}
\end{figure}

Table 2 demonstrates the improvement performance of network, as a new component is added to the architecture, such as visual information, multi-layer feature fusion strategy, and finally GAN model. The VSEGAN outperforms SEGAN, which is an evidence that visual information significantly improves the performance of speech enhancement system. What is more, the comparison between VSEGAN and baseline illustrates that GAN model for visual speech enhancement is more robust than G-only model. Hence the performance improvement from SEGAN to VSEGAN is primarily for two reason: 1) using visual information, and 2) using GAN model. Figure~\ref{fig:spe} shows the visualization of baseline system enhancement, Generator-only enhancement, and VSEGAN enhancement, which most obvious details of spectrum distinction are framed by dotted box.\footnote{Speech samples are available at: \scriptsize{\href{https://XinmengXu.github.io/AVSE/VSEGAN}{\texttt{https://XinmengXu.github.\\io/AVSE/VSEGAN}}}}

For further investigating the superiority of proposed method, the performance of VSEGAN has also compared to the following recent audio-visual speech enhancement approaches on GRID dataset:
\begin{itemize}
    \item \textbf{Looking-to-Listen model} \cite{ref42}: A speaker independent audio-visual speech separation model.
    \item  \textbf{Online Visual Augmented (OVA) model} \cite{wang2020robust}: A late fusion based visual speech enhancement model, which involves the audio-based component, visual-based component and the augmentation component.
    \item \textbf{AV(SE)$^2$ model} \cite{iuzzolino2020av}: An audio-visual squeeze-excite speech enhancement model.
\end{itemize}

Table 3 shows that the VSEGAN produces state-of-the-art results in terms of PESQ and STOI score by comparing against four recent proposed methods that use DNNs to perform end-to-end visual speech enhancement. Results for competing methods are taken from the corresponding papers and the missing entries in the table indicate that the metric is not reported in the reference paper. Although the competing results are for reference only, the VSEGAN has better performance than state-of-the-art results on the GRID dataset.

\section{Conclusions}
This paper proposed an end-to-end visual speech enhancement method has been implemented within the generative adversarial framework. The model adopts multi-layer feature fusion convolution network structure, which provides a better training behavior, as the gradient can flow deeper through the whole structure. According to the experiment results, the performance of speech enhancement system has significantly improves by involving of visual information, and visual speech enhancement using GAN performs better quality of enhanced speech than several state-of-the-art models.
\vfill\pagebreak

\bibliographystyle{IEEEbib}
\bibliography{strings,refs}

\begin{thebibliography}{10}

\bibitem{loizou2013speech}
Philipos~C Loizou,
\newblock {\em Speech enhancement: theory and practice},
\newblock CRC press, 2013.

\bibitem{martin1994spectral}
Rainer Martin,
\newblock ``Spectral subtraction based on minimum statistics,''
\newblock {\em power}, vol. 6, no. 8, 1994.

\bibitem{ephraim1995signal}
Yariv Ephraim and Harry~L Van~Trees,
\newblock ``A signal subspace approach for speech enhancement,''
\newblock {\em IEEE Transactions on speech and audio processing}, vol. 3, no.
  4, pp. 251--266, 1995.

\bibitem{lim1978all}
Jae Lim and Alan Oppenheim,
\newblock ``All-pole modeling of degraded speech,''
\newblock {\em IEEE Transactions on Acoustics, Speech, and Signal Processing},
  vol. 26, no. 3, pp. 197--210, 1978.

\bibitem{dendrinos1991speech}
Markos Dendrinos, Stelios Bakamidis, and George Carayannis,
\newblock ``Speech enhancement from noise: A regenerative approach,''
\newblock {\em Speech Communication}, vol. 10, no. 1, pp. 45--57, 1991.

\bibitem{sumby1954visual}
William~H Sumby and Irwin Pollack,
\newblock ``Visual contribution to speech intelligibility in noise,''
\newblock {\em The journal of the acoustical society of america}, vol. 26, no.
  2, pp. 212--215, 1954.

\bibitem{wang2005video}
Wenwu Wang, Darren Cosker, Yulia Hicks, S~Saneit, and Jonathon Chambers,
\newblock ``Video assisted speech source separation,''
\newblock in {\em Proceedings.(ICASSP'05). IEEE International Conference on
  Acoustics, Speech, and Signal Processing, 2005.} IEEE, 2005, vol.~5, pp.
  v--425.

\bibitem{hou2016audio}
Jen-Cheng Hou, Syu-Siang Wang, Ying-Hui Lai, Jen-Chun Lin, Yu~Tsao, Hsiu-Wen
  Chang, and Hsin-Min Wang,
\newblock ``Audio-visual speech enhancement using deep neural networks,''
\newblock in {\em 2016 Asia-Pacific Signal and Information Processing
  Association Annual Summit and Conference (APSIPA)}. IEEE, 2016, pp. 1--6.

\bibitem{gabbay2018seeing}
Aviv Gabbay, Ariel Ephrat, Tavi Halperin, and Shmuel Peleg,
\newblock ``Seeing through noise: Visually driven speaker separation and
  enhancement,''
\newblock in {\em IEEE International Conference on Acoustics, Speech and Signal
  Processing (ICASSP)}. IEEE, 2018, pp. 3051--3055.

\bibitem{inproceedings}
Santiago Pascual, Antonio Bonafonte, and Joan Serrà,
\newblock ``{SEGAN}: Speech enhancement generative adversarial network,''
\newblock in {\em Interspeech 2017}, 2017, pp. 3642--3646.

\bibitem{mirza2014conditional}
Mehdi Mirza and Simon Osindero,
\newblock ``Conditional generative adversarial nets,''
\newblock {\em Computer ence}, pp. 2672--2680, 2014.

\bibitem{aaa}
Alec Radford, Luke Metz, and Soumith Chintala,
\newblock ``Unsupervised representation learning with deep convolutional
  generative adversarial networks,''
\newblock 11 2016.

\bibitem{mao2017least}
Xudong Mao, Qing Li, Haoran Xie, Raymond~YK Lau, Zhen Wang, and Stephen
  Paul~Smolley,
\newblock ``Least squares generative adversarial networks,''
\newblock in {\em Proceedings of the IEEE international conference on computer
  vision}, 2017, pp. 2794--2802.

\bibitem{baby2019sergan}
Deepak Baby and Sarah Verhulst,
\newblock ``Sergan: Speech enhancement using relativistic generative
  adversarial networks with gradient penalty,''
\newblock in {\em ICASSP 2019-2019 IEEE International Conference on Acoustics,
  Speech and Signal Processing (ICASSP)}. IEEE, 2019, pp. 106--110.

\bibitem{isola2017image}
Phillip Isola, Jun-Yan Zhu, Tinghui Zhou, and Alexei~A Efros,
\newblock ``Image-to-image translation with conditional adversarial networks,''
\newblock in {\em Proceedings of the IEEE conference on computer vision and
  pattern recognition}, 2017, pp. 1125--1134.

\bibitem{pandey2018adversarial}
Ashutosh Pandey and Deliang Wang,
\newblock ``On adversarial training and loss functions for speech
  enhancement,''
\newblock in {\em 2018 IEEE International Conference on Acoustics, Speech and
  Signal Processing (ICASSP)}. IEEE, 2018, pp. 5414--5418.

\bibitem{cooke2006audio}
Martin Cooke, Jon Barker, Stuart Cunningham, and Xu~Shao,
\newblock ``An audio-visual corpus for speech perception and automatic speech
  recognition,''
\newblock {\em The Journal of the Acoustical Society of America}, vol. 120, no.
  5, pp. 2421--2424, 2006.

\bibitem{c}
Naomi Harte and Eoin Gillen,
\newblock ``{TCD-TIMIT}: An audio-visual corpus of continuous speech,''
\newblock {\em IEEE Transactions on Multimedia}, vol. 17, no. 5, pp. 603--615,
  2015.

\bibitem{ref39}
Aviv Gabbay, Asaph Shamir, and Shmuel Peleg,
\newblock ``Visual speech enhancement,''
\newblock {\em Interspeech}, pp. 1170--1174, 2018.

\bibitem{ref42}
Ariel Ephrat, Inbar Mosseri, Oran Lang, Tali Dekel, Kevin Wilson, Avinatan
  Hassidim, William~T Freeman, and Michael Rubinstein,
\newblock ``Looking to listen at the cocktail party: A speaker-independent
  audio-visual model for speech separation,''
\newblock {\em ACM Transactions on Graphics}, 2018.

\bibitem{wang2020robust}
Wupeng Wang, Chao Xing, Dong Wang, Xiao Chen, and Fengyu Sun,
\newblock ``A robust audio-visual speech enhancement model,''
\newblock in {\em ICASSP 2020-2020 IEEE International Conference on Acoustics,
  Speech and Signal Processing (ICASSP)}. IEEE, 2020, pp. 7529--7533.

\bibitem{iuzzolino2020av}
Michael~L Iuzzolino and Kazuhito Koishida,
\newblock ``{AV(SE)$^2$}: Audio-visual squeeze-excite speech enhancement,''
\newblock in {\em ICASSP 2020-2020 IEEE International Conference on Acoustics,
  Speech and Signal Processing (ICASSP)}. IEEE, 2020, pp. 7539--7543.

\end{thebibliography}

\end{document}